\documentclass[12pt]{article}
\usepackage{amsmath}
\usepackage{bm}
\usepackage{amssymb}
\usepackage{graphicx,xcolor}
\usepackage{epstopdf}
\usepackage{bbm}
\usepackage{mathrsfs}

\newcommand{\qed}{ $\blacksquare$} \newcommand{\nn}{\nonumber}
\newcommand{\R}{{\mathbb R}}

\newcommand{\N}{{\mathbb N}}
\renewcommand{\a}{\alpha}

\renewcommand{\d}{\delta}

\newcommand{\pa}{\partial}

\newcommand{\D}{\Delta}

\def\RRR{{\mathbb R}}

\def\TT{\mathtt{T}}

\def\be{\begin{equation}}
\def\ee{\end{equation}}
\def\bea{\begin{eqnarray}}
\def\eea{\end{eqnarray}}

\def\nn{\nonumber}

\def\d{\delta}

\title {\LARGE Propagation of chaos for topological interactions }

\date{}

\begin{document}

\maketitle

\begin{center}

{\large P. Degond $^{1}$ and M. Pulvirenti $^{2}$}

\vspace{0.5cm}
{$1.$\scshape {\small \ Department of Mathematics
Imperial College London
London SW7 2AZ, UK \\	
pdegond@imperial.ac.uk  \\ \smallskip}
{$2.$ International Research Center on the Mathematics and Mechanics of Complex Systems MeMoCS, University of L'Aquila, Italy\\
pulviren@mat.uniroma1.it}}

\end{center}

\vspace{1.5cm}
\noindent


\begin{abstract}

We consider a $N$-particle model describing an alignment mechanism due to a topological interaction among the agents. We show that the kinetic equation, expected to hold in the mean-field limit $N \to \infty$, as following from the previous analysis in Ref. \cite
{Blanchet_Degond_JSP16} can be rigorously derived. This means that the statistical independence (propagation of chaos) is indeed recovered in the limit, provided it is assumed at time zero. 

\end{abstract}

\medskip



\bigskip

{\bf Aknowledgments} PD acknowledges support by the Engineering and Physical Sciences
Research Council (EPSRC) under grants no. EP/M006883/1 and EP/P013651/1 by the Royal
Society and the Wolfson Foundation through a Royal Society Wolfson
Research Merit Award no. WM130048 and by the National Science
Foundation (NSF) under grant no. RNMS11-07444 (KI-Net). PD is on leave
from CNRS, Institut de Math\'ematiques de Toulouse, France.

\medskip

{\bf Data statement} : no new data were collected in the course of this research.

\large
\section{Introduction}

Propagation of chaos is a fundamental property in Kinetic Theory: it allows to pass from a $N$-particle description, which is usually intractable due to the huge number of particles to handle, to a single partial differential equation. Originally it refers to deterministic particle systems and it has been introduced by Boltzmann in the formal derivation of his famous equation. From the mathematical side we address to the well known paper by Lanford  \cite {Lanford} (see also  \cite{Bodi}, \cite{CIP}, \cite {GST}, \cite{IP86}, \cite{IP89}, \cite {PSS}, \cite {PS}, \cite{Spohn1}, \cite {Uk01} for  subsequent progresses) where the  validity of the Boltzmann equation has been proved for a short time interval. On the other hand other stochastic processes have been introduced to derive the Boltzmann equation and the most famous model is Kac's model \cite {Kac}, \cite{Kac2}. See also \cite {MM} and \cite {PPS} for recent developments. Similar models of interest for the numerics have also been studied for instance in  \cite{LP} \cite{PWZ} \cite{RW}.  Nowadays the methodology and techniques of Kinetic Theory have been applied also to mean-field limits of particle models in which interactions are averages of binary interactions and which, at the kinetic level, give rise to non linear Vlasov (in the deterministic case) or Fokker-Planck (in the stochastic case) equations, see e.g. \cite{NeunWick}  \cite{Bolley_etal}, \cite{Dobru}, \cite{Hauray_Jabin}, \cite{Jabin_Wang}, \cite{Lions_Sznitman}, \cite{Sznitman}. For recent approaches to propagation of chaos see \cite{MMW}.

In most mean-field models, binary interactions are weighted by a function of the relative distance between the two particles. However, recent observations~\cite{Ballerini_etal_PNAS08, Cavagna_etal_M3AS10} have shown that interactions between animals in nature are weighted by a function of their rank, irrespective of the relative distance, meaning that the interaction probability of an individual with its $k$-th nearest neighbor is the same whether this individual is close or far. This new type of interaction has been called ``topological'', by contrast to the usual ``metric'' interaction which is a function of the subjects' relative distance. Numerical simulations of particle systems undergoing topological interactions seem to support the observations \cite{Bode_etal_Interface10, Camperi_etal_InterfaceFocus12, Ginelli_Chate_PRL10}. In the recent past, the literature on the applications of topological interactions to flocking has grown exponentially \cite{Hemelrijk_Hildenbrandt_PlosOne11}, \cite{Jian_etal_ChinPhysB10}, \cite{Niizato_etal_Biosystems14}, \cite{Shang_Bouffanais_EurPhysJB14}. On the mathematical side, flocking under topological interactions has been studied in \cite{Haskovec_PhysD13, Martin_SystContLett14, Shang_Bouffanais_SciRep14, Wang_Chen_Chaos16}. In \cite{Haskovec_PhysD13} mean-field kinetic and fluid models for topological mean-field interactions are formally derived. Recently, \cite{Blanchet_Degond_JSP16} and \cite{Blanchet_Degond_JSP17} have formally derived kinetic models for jump processes ruled by topological interactions. In the former, the number of particles interacting with a given particle is unbounded in the large particle number limit, while in the latter, particles only interact with a fixed finite number of closest neighbors. In the large particle number limit, the former gives rise to an interaction operator in integral form, while the latter provides a diffusion-like interaction operator. 

The goal of this paper is to give a rigorous proof of convergence for the jump process of \cite{Blanchet_Degond_JSP16} in the limit of the number of particles tending to infinity, i.e. to prove that propagation of chaos holds for this system in this limit,  providing a rigorous derivation of the kinetic equation. 

Here new difficulties arise. Indeed in usual metric models particles interact through two-body interactions which are averaged through weights that depend on the distance between the two interacting particles. This structure reflects in the system satisfied by the hierarchy of joint probability distributions (also known as the BBGKY hierarchy): the evolution of the $s$-th marginal only depends on the $s+1$-th marginal. This structure is lost with topological interactions, as the rank of a particle neighbor depends on all the other particles. 
Now the study of the hierarchy usually describing the time evolution of the marginals is not possible anymore: the time evolution of the $s$-particle marginal depends on the full $N$-particle probability measure. Therefore, to prove propagation of chaos,  we are facing new, previously unmet, problems.

Obviously the hierarchical approach is not the only possible one. For instance we quote \cite{GM97} where Kac's model has been treated by a coupling technique, yielding, by the way, optimal estimates. Such a technique is not easy to apply to the present context.    Our strategy is different. We  assume the function that weights the interaction strength with the various partners to be real analytic. For such a kind of interactions we can establish a new hierarchy for which the time evolution of the $j$-particle marginal $f_j$ is expressed in terms of an infinite sequence of marginals $f_m$ with $m>j$, with decreasing weight.

\section{The model}  \label{sec:1}
\setcounter{equation}{0}    
\def\theequation{2.\arabic{equation}}

Here, we recall the setting of \cite{Blanchet_Degond_JSP16}. We consider a $N$-particle system in $\RRR^d$, $d=1,2,3 \dots$ ( or in $\TT^d$ the $d$-dimensional torus).
Each particle, say particle $i$, has a position $x_i$ and velocity $v_i$. The configuration of the system is denoted by
$$
Z_N=\{z_i\}_{i=1}^N=\{x_i,v_i \}_{i=1}^N= (X_N,V_N).
$$ 

Given the particle $i$, we order the remaining particles $j_1,j_2, \cdots j_{N-1}$ according their distance from $i$, namely by the following relation
$$
| x_i -x_{j_s} | \leq | x_i -x_{j_{s+1}} |, \qquad s=1,2 \cdots N-1.
$$
The rank (with respect to  $i$) of particle $k=j_s$ is $s$.  The rank is denoted by $R(i,k)$.

The normalized rank is defined as
$$
r(i,k) = \frac {R(i,k)}{N-1} \in \{ \frac 1{N-1}, \frac 2{N-1} , \dots \}.
$$

Next we introduce a (smooth) function
$$
K: [0,1]\to \RRR^+  \qquad \mbox {s.t.} \,\, \int_0^1 K(r) dr =1 , 
$$
 and the following quantities
\be
\label{prob}
\pi_{i,j} = \frac {K(r(i,j))} {\sum_s K( \frac s {N-1} )}.
\ee
Clearly 
$$
\sum_{j}^{} \pi_{i,j}=1 .
$$

We are now in the right position to introduce a stochastic process describing alignment via a topological interaction.
The particles go freely, namely following the trajectory  $ x_i+v_i t$. At some random time dictated by a Poisson process of intensity $N$,  a particle (say $i$) is chosen with probability $\frac 1N$ and a partner particle, say $j$, with probability $\pi_{i,j}$. Then  the transition
$$
(v_i,v_j) \to (v_j, v_j).
$$
is performed. After that the system goes freely with the new velocities and so on. 

The process is fully described by the continuous-time Markov  generator given, for any $\Phi \in C^1_b (\R^{2dN})$ by
\bea
\label{generator}
&& L_N \Phi ( x_1,v_1, \cdots x_N,v_N)  =   \sum\limits_{i=1}^{N} v_i \cdot \nabla_{x_i} \Phi ( x_1,v_1, \cdots x_N,v_N)+  \\
&& \sum\limits _{i=1}^N \sum\limits_{\substack {1\leq j \leq  N \\ i\neq j}} \pi_{i,j}
\big[ \Phi ( x_1,v_1, \cdots x_i v_j \cdots  x_j,v_j \cdots x_N,v_N) - 
 \Phi ( x_1,v_1, \cdots x_N,v_N)  \big]. \nn
\eea

Note that $\pi_{i,j}=\pi_{i,j}^N$ depends not only on $N$ but also on the whole configuration $Z_N$. 

The law of the process $W^N(Z_N;t)$ is driven by the following evolution equation
\bea
\label{master}
&&\pa_t  \int W^N(t) \Phi = \int W^N (t) \sum\limits_{i=1}^{N} v_i \cdot \nabla_{x_i} \Phi+ 
\int W^N (t) 
\sum\limits _{i=1}^N  
  \sum\limits_{\substack {1\leq j \leq  N \\ i\neq j}} \pi_{i,j} \nn  \\
&& 
\big[ \Phi (  x_1,v_1, \cdots x_i v_j \cdots  x_j,v_j \cdots x_N,v_N) 
  -\Phi ( x_1,v_1, \cdots x_N,v_N)  \big] ,  
\eea
for any test function $\Phi$.

We assume that the initial measure $W^N_0=W^N(0)$ factorizes, namely $W^N_0= f_0^{\otimes N} $ where $f_0$ is the initial datum for the limiting kinetic equation we are going to establish. Note also that $W^N(Z_N;t)$, for $t \geq 0$,  is symmetric in the exchange of particles.

The strong form of  Eq. \eqref{master} is
\be
\label{masters1}
(\pa_t +\sum\limits_{i=1}^{N} v_i \cdot \nabla_{x_i} )W^N(t) =-N W^N(t)+{\cal L}_NW^N(t), 
\ee 
where
\be
\label{masters2}
{\cal L}_N W^N(X_N,V_N, t)= \sum\limits _{i=1}^N  \sum\limits_{\substack {1\leq j \leq  N \\ i\neq j}} \int du \, \pi_{i,j} \,
 W^N (X_N, V_N^{(i)}(u)) \delta (v_i-v_j) . 
\ee
Here $V_N^{(i)}(u)= (v_1 \cdots v_{i-1},u,v_{i+1} \cdots v_N)$  \, if \,  $V_N = (v_1 \cdots v_{i-1},v_i,v_{i+1} \cdots v_N)$.

\section{Kinetic description}  \label{sec:3}
\setcounter{equation}{0}    
\def\theequation{3.\arabic{equation}}

 Here we present a heuristic derivation of the kinetic equation we expect to be valid in the limit $N \to \infty$. This derivation is slightly simpler than in \cite{Blanchet_Degond_JSP16}. 

We first compute explicitly the transition probability $\pi_{i,j}$. In general:
$$
r(i,j) = \frac 1 {N-1} \sum\limits_{\substack{ 1 \leq k \leq N \\ k \neq i }} \chi_{B(x_i, |x_i-x_j|)} (x_k) , 
$$
where  $\chi_{B(x_i, |x_i-x_j|)}$ is the characteristic function of the ball $\{y \, \, | \, \,  |x_i-y| \leq |x_i-x_j| \}$.
Moreover, recalling that $\int K=1$,
\bea
\label{K}
\sum_s K( \frac s {N-1}) && = (N-1) \big( 1-\int_0^1 K(x)dx+ \frac 1{N-1}\sum_s K( \frac s {N-1})  \big) \nn \\
&& =(N-1)(1-e_K(N)) , \nn 
\eea
where the last identity defines $e_K(N)$. Note that $e_K$ measures the difference between the integral and the Riemann sum of $K$.

Clearly
\be
\label{este}
| e_K(N)| \leq \| K' \|_{L^{\infty}} \frac 1 {N-1}.
\ee

Therefore by \eqref{prob}
\be
\label{prob1}
\pi_{i,j} = \a_N K( \frac 1 {N-1} \sum_{k \not = i} \chi_{B(x_i, |x_i-x_j|)} (x_k) ) , 
\ee
where 
\be
\label{alpha}
\a_N= \frac 1 { (N-1) (1-e_K(N) )}.
\ee

\bigskip

Setting $\Phi(Z_N)=\varphi (z_1)$ in \eqref {master},  we obtain
\be
\label{kin1}
\pa_t \int f^N_1 \varphi = \int f^N_1 v \cdot \nabla_{x} \varphi -\int f^N_1 \varphi  
+\int W^N \sum_{j\neq 1} \pi_{i,j} \varphi (x_1, v_j).
\ee
Here $f^N_1$ denotes the one-particle marginal of the measure $W^N$. We recall that the $s$-particle marginals are defined by
$$
f^N_s (Z_s) =\int W^N (Z_s, z_{s+1} \cdots z_N) dz_{s+1} \cdots dz_N, \qquad s=1,2 \cdots N, 
$$
and are the distribution of the first $s$ particles (or of any group of $s$ tagged  particles).

In order to describe the system in terms of a single kinetic equation, we expect that chaos propagates.  Actually since $W^N$ is initially factorizing, although the dynamics creates correlations, we hope that, due to the weakness of the interaction, factorization still holds approximately also at any positive time $t$, namely
$$
f^N_s \approx f_1 ^{\otimes s}
$$
for any fixed integer $s$.
In this case the strong law of large numbers does hold, that is for almost all i.i.d. variables $ \{ z_i(0) \}$ distributed according to $f_1(0)=f_0$, the random measure
$$
\frac 1N \sum_j \delta (z-z_j (t) )  
$$
approximates  weakly $ f^N_1(z,t) $.
Then
\bea
\pi_{i,j} \approx  \frac 1{N-1} K( \frac 1{N-1}  \sum_{k \not = i} \chi_{B(x_i, |x_i-x_j|)} (x_k) ) \nn \\
 \approx \frac 1{N-1} K( M_{\rho} ( x_i, |x_i-x_j|)), 
\eea
where
$$
M_{\rho} (x,R)= \int_{B(x,R)} \rho( y ) dy,
$$
and where $\rho (x)  =\int dv f^N_1 (x,v) $ is the spatial density and  $B(x,R)$ is the ball of center $x$ and radius $R$.

In conclusion we expect that, by \eqref {kin1}, using the symmetry of $W^N$,  $f^N_1 \to f$ and $f^N_2 \to f^{\otimes 2} $ in the limit $N\to \infty$, where $f$ solves
\be
\label{kin2}
\pa_t \int f \varphi = \int f v \cdot \nabla_{x} \varphi -\int f \varphi  
+\int f(z_1) f(z_2)  \varphi (x_1, v_2) K( M_{\rho} ( x_1, |x_1-x_2|)) , 
\ee
or, in  strong form,
\be
\label{eqkin}
(\pa_t + v \cdot \nabla_x )f= -f + \rho(x) \int dy  K (M_{\rho} ( x, |x-y|) ) \,f(y,v), 
\ee
which is the equation we want to derive rigorously.

As regards existence and uniqueness of the solutions to Eq.  \eqref{eqkin} we can apply the Banach fixed point theorem in $L^1(x,v)$ to find a unique solution for  \eqref{eqkin} in mild form, for a short time interval, provided that $K$ has bounded derivative in $[0,1]$. The global solution is recovered by the conservation of the
$L^1(x,v)$ norm.  The method is classical  and we leave the details to the reader.

\section{Hierarchies}  
\setcounter{equation}{0}    
\def\theequation{4.\arabic{equation}}

We assume the function $K$  to be expressible in terms of a power series, 
\be
\label{pol}
K(x)= \sum_{m=0}^\infty  a_m  x^m , \qquad x \in [0,1].
\ee
for some sequence of coefficients $a_m$. The normalization condition gives the constraint $a_0+\sum_{m \geq 1}^M \frac 1 {m+1} a_m=1$. Note that the coefficients $a_m$ are not necessarily positive.

We further assume that
\be
\label{hypo}
A:= \sum_{m=0}^\infty  |a_m | 8^m < +\infty
\ee

\medskip

{\bf Remark}

{\it An example of a function $K$ satisfying the above hypotheses is, for $x \in (0,1)$:
$$
K(x) = \frac {e^{1-x} -1}{e-2} =  \frac 1{e-2} (e-1+ e\sum_{r \geq 1} \frac { (-1)^r x^r}{r!} ).
$$ 
}

\bigskip

To outline the behavior of the $s$- particle marginal $f^N_s$ we integrate \eqref{masters1} with respect to the last $N-s$ variables and compute preliminarly
$$
\sum\limits _{i=s+1}^N  \sum\limits_{\substack {1\leq j \leq  N \\ i\neq j}} \int du \pi_{i,j} \,\, W^N (X_N, V_N^{(i)}(u)) \delta (v_i-v_j)dz_{s+1} \cdots dz_N =(N-s) f_s^N (X_s,V_s), 
$$
since the variable $z_i$ is integrated. Therefore 
\bea
\label{hie1}
&&\hspace{-1cm} 
(\pa_t +\sum_{i=1}^{s} v_i \cdot \nabla_{x_i} )f_s^N(t) =-s f_s^N (t)+ E^1_s(t) +\\
&& \hspace{1cm} (N-s)  \sum\limits _{i=1}^s  \int dz_{s+1} \cdots dz_N \,  \pi_{i,s+1} 
 W^N (X_N, V_N^{(i,s+1)}; t),   \nn
\eea 
where
$$
V_N^{(i,s+1)}= \{ v_1 \cdots v_{i-1}, v_{s+1}, v_{i+1} \cdots v_s, v_i ,v_{s+2} \cdots v_N \},
$$
namely the velocities of particles $i$ and $s+1$ exchange their positions in the sequence $V_N=\{ v_1 \cdots  v_N \}$, and
\bea
\label{error1}
E^1_s(t)=
\sum\limits _{i=1}^s  \sum\limits_{\substack {1\leq  j \leq  s \\ i\neq j}} \int du \, dz_{s+1} \cdots dz_N \,  \pi_{i,j} 
 W^N (X_N, V_N^{(i)}(u);t) \delta (v_i-v_j).
\eea
We expect $E^1_s$ to be $O(\frac {s^2}N)$  since $\pi_{i,j} =O(\frac 1N)$ (see \eqref{prob1} and \eqref{alpha}). This is the first error term entering in the present analysis. A precise estimate of this term is forthcoming. Note also that we used the symmetry to deduce the last term in the right hand side of \eqref{hie1}.

Next, setting $\chi_{i,j}=\chi_{B(x_i, |x_i-x_j|)}$, we have from \eqref{prob1} and \eqref{pol}
\be
\label{pi}
\pi_{i,j} =  \a_N \sum_{r=0}^\infty  a_r  \frac 1{(N-1)^r}  \sum_{(k_1,k_2 \cdots k_r) \in (\{1,N \} \setminus \{i\})^r} \,\,\,\chi_{i,j}(x_{k_1}) \dots \chi_{i,j} (x_{k_r}).  
\ee
Inserting this quantity into the last term of \eqref{hie1}, we obtain
\bea
\label{hie2}
(\pa_t +\sum\limits_{i=1}^{s} &&v_i \cdot \nabla_{x_i} )f_s^N(t)=-s f_s^N (t)+ E^1_s(t) +E^2_s \\
&& +(N-s) \a_N    \sum_{r=0}^\infty  a_r \, C^N_{s,s+r+1} \, f^N_{s+r+1} , 
 \nn
\eea 
where $C^N_{s,s+r+1}: L^1(\R^{2d(s+r+1)}) \to L^1( \R^{2ds}  )$ is a linear operator defined by
\bea
\label{C}
&& C^N_{s,s+r+1}g_{s+r+1}(X_s,V_s)= \frac {(N-s-1)\dots (N-s-r)}{(N-1)^r}   \sum\limits_{i=1}^{s} \\
&& \int dz_{s+1} \cdots dz_{s+r+1} \,\, \chi_{i,s+1}(x_{s+2}) \dots \chi_{i,s+1} (x_{s+r+1}) \,\,   
g_{s+r+1}(X_{s+r+1}, V^{(i,s+1)} _{s+r+1}). \nn
\eea

The form \eqref{C}   of the operator $C^N_{s,s+r+1}$ comes from  considering in the sum  $\sum_{k_1,k_2 \cdots k_r}$ 
in \eqref{pi},  only the contributions given by
$$
\sum_{\substack {k_1 \neq k_2 \cdots \neq k_r \\ k_m >s+1; \,\,m=1 \dots r}} , 
$$
namely all the $k_m$ are different and larger than $s+1$. Clearly we also used the symmetry. The term $E^2_s$ is what remains, namely
\bea
\label{error2}
E^2_s(Z_s) && = (N-s) \a_N \sum_{i=1}^s\sum_{r=0}^\infty a_r  (\frac 1 {N-1} )^r \sum^*_{k_1,k_2 \cdots k_r} \\
&& \int dz_{s+1} \cdots dz_N
\chi_{i,s+1}(x_{k_1}) \dots \chi_{i,s+1} (x_{k_r}) W^N (Z_s, z_{s+1} \cdots z_N;t) ,  \nn
\eea
with
$$
\sum^*_{k_1,k_2 \cdots k_r}=
\sum_{\substack {k_1, k_2 \cdots k_r \\ k_m  \neq i, m=1 \dots r}}-
\sum_{\substack {k_1 \neq k_2 \cdots \neq k_r \\ k_m >s+1, m=1 \dots r}}.
$$

Again we expect that $E^2_s$ is negligible in the limit as we shall see in a moment.

Note that for $s=N$  \eqref {hie2} becomes identical to Eq. \eqref{master} as the last two terms are equal to zero. 
We will also use the convention that $f^N_s(t)=0$ if $s>N$.

We have to compare Eq. \eqref{hie2} with a similar hierarchy satisfied by the sequence of marginals $f_j(t)=f^{\otimes j} (t)$,  where $f$ solves the kinetic equation. Such a hierarchy is easily recovered. Indeed
coming back to the kinetic equation \eqref {eqkin} we observe that, by virtue of \eqref{pol}
\bea
&& \hspace{-1cm}
K (M_\rho( x_i, |x_i-x_{s+1}|) )= \sum_r a_r \\
&& \hspace{1cm} \int dz_{s+2} \cdots dz_{s+r+1} \chi_{i,s+1}(x_{s+2}) \dots \chi_{i,s+1} (x_{k_{s+r+1}}) f^{\otimes r} (z_{s+2} \cdots 
z_{s+r+1}) , \nn
\eea
and \eqref {eqkin}  becomes (recalling that $z_1 =(x_1,v_1)$):
\bea
\label{eqkin2}
&& \hspace{-1cm}
(\pa_t +v_1 \cdot \nabla_{x_1} )f (z_1,t)+f(z_1,t) = \sum_{r=0}^\infty a_r   \int dz_2 \dots  \int dz_{2+r}  \chi_{1,2}(x_3)  \cdots  \chi_{1,2}(x_{2+r}) \nn \\ 
&& \hspace{1cm} \cdot f(x_1,v_2;t) f(x_2,v_1;t)
f^{\otimes r} (z_3 \cdots z_{2+r} ;t ).
\eea

As a consequence an easy computation shows that $f_s=f^{\otimes s} $ solves

\bea
\label{hielim}
(\pa_t +\sum\limits_{i=1}^{s} &&v_i \cdot \nabla_{x_i} )f_s(t)=-s f_s (t)+  \\
&& + \sum_{r=0}^\infty  a_r \, C_{s,s+r+1} \, f_{s+r+1} , 
\nn
\eea 
where 
\bea
\label{Clim}
&& C_{s,s+r+1}f_{s+r+1}(X_s,V_s)=  \\
&&  \sum\limits_{i=1}^{s} \int dz_{s+1} \cdots dz_{s+r+1} \chi_{i,s+1}(x_{s+2}) \dots \chi_{i,s+1} (x_{k_{s+r+1}})    
f_{s+r+1}(X_{s+r+1}, V^{(i,s+1)} _{s+r+1}). \nn
\eea

In view of the comparison of  $f^N_s$ with $f_s$ we rewrite \eqref{hie2} as
\bea
\label{hie3}
(\pa_t +\sum\limits_{i=1}^{s} &&v_i \cdot \nabla_{x_i} )f_s^N(t)=-s f_s^N (t)+ E_s(t) \\
&& +    \sum_{r=0}^\infty  a_r \, C_{s,s+r+1} \, f^N_{s+r+1} , 
 \nn
\eea 
where 
\be
\label{E}
E_s=E^1_s(t) +E^2_s(t)+E^3_s(t) 
\ee
and
\be
\label{E3}
E^3_s(t)= (N-s) \a_N    \sum_{r=0}^\infty  a_r \, C^N_{s,s+r+1} \, f^N_{s+r+1}-
 \sum_{r=0}^\infty a_r \, C_{s,s+r+1} \, f^N_{s+r+1}
\ee

The initial conditions for \eqref {hie3} and \eqref {hielim} are
$$
f^N_s (0) = f_0^{\otimes s} {\bf 1}_{\{ s \leq N \}}
$$
where $ {\bf 1}_{\{ s \leq N \}}$ is the indicator of the set  $\{ s \leq N \} $
and
$$
f_s (0) = f_0^{\otimes s} 
$$
respectively. Here $f_0 \in L^1$ is the initial datum of the kinetic equation.

\section{Estimates of the error term}  \label{sec:2}
\setcounter{equation}{0}    
\def\theequation{5.\arabic{equation}}

In this section we establish some  estimates of the error term $E_s$ appearing in eEq. \eqref {hie3}.

We observe preliminarily that, by the particular form of the function $K$ given by \eqref{pol}, 
we have,  $\| K' \|_{L^\infty} \leq A$ and, using \eqref {este},
\be
\label{eK}
 |e_K (N)| \leq \frac {A}{N-1}.
\ee
Therefore 
\be
\label{esta}
\a_N = \frac 1 { (N-1) (1-e_K(N) )} \leq \frac {4 e^ {|e_K(N)|}}{N-1} \leq  \frac {4 e^ {\frac {A} {N-1} }}{N-1}, 
\ee
for $N>2A+1$.
This follows by the obvious  inequality
$$
\frac{1}{1-x} \leq 4 e^{x} 
$$
valid for $x \in (0, \frac 12)$

As a consequence, by \eqref{prob1} and from the fact that $ \| K \|_{L^\infty} \leq A$,
\be
\label{estpi}
\pi_{i,j} \leq \a_N A \leq \frac {4Ae^ {\frac {A} {N-1} }}{N-1}.
\ee

The operators $C^N$ and $C$ are easily estimated:
\be
\label{boundC}
\max (\| C^N_{s,s+r+1}g_{s+r+1}\|_{L^1}, \| C_{s,s+r+1}g_{s+r+1}\|_{L^1}) \leq s \| g_{s+r+1}\|_{L^1} ,
\ee
due to the fact that $\chi \leq 1$ and that the prefactor in formula \eqref{C} is less than unity.

As regards the error terms \eqref{error1} we have, by \eqref{estpi}
\be
\label{esterr1}
\| E_s^1 (t) \|_{L^1} \leq  s^2  \,\, \frac {4Ae^ {\frac {A} {N-1} }}{N-1}.
\ee
Strictly speaking here we make a notational abuse. $E^1$ is a measure so that $\| E_s^1 (t) \|_{L^1}$ has to be understood as the total variation norm. In other words $\| \mu \|_{L^1}$  is the $L^1$ norm of the densities whenever $\mu$ is absolutely continuous. Otherwise it is the total variation.

Moreover by \eqref {error2} and \eqref{esta}
\be
\| E_s^2 (t) \|_{L^1} \leq  4e^ {\frac {A} {N-1}}  (\frac { N-s} {N-1} ) \sum_{i=1}^s\sum_{r=0}^\infty a_r  (\frac 1 {N-1} )^r \sum^*_{k_1,k_2 \cdots k_r} 1.
\ee
But
$$
 \sum^*_{k_1,k_2 \cdots k_r}1\leq  \sum^{**}_{k_1,k_2 \cdots k_r}1+\sum^{***}_{k_1,k_2 \cdots k_r}1, 
$$
where $\sum^{**}_{k_1,k_2 \cdots k_r}1$ means that  $k_m \leq s+1$ for  at least one $m=1,2 \cdots r$, while
$\sum^{***}_{k_1,k_2 \cdots k_r}$ means that all the $k_m$ are larger than $s+1$ but $k_\ell=k_m$ for at least one couple $\ell,  m$ in $1,2 \cdots r$. 

Moreover, denoting by $\ell$ the number of indices $m$ for which $k_m \leq s+1$, we have
$$
 \sum^{**}_{k_1,k_2 \cdots k_r}1=\sum_{\ell=1}^r \binom r \ell s^{\ell} (N-s-1)^{r-\ell}=(N-1)^r-(N-s-1)^r \leq rs\, (N-1)^{r-1},
$$

where in the last step we used the Taylor expansion of the function $x^r$ with initial point $N-s-1$.

Furthermore
$$
\sum^{***}_{k_1,k_2 \cdots k_r}1 \leq \frac {r (r-1)}2 (N-s-1)^{r-1}.
$$

Therefore

\bea
\| E_s^2 (t) \|_{L^1} \leq && 4e^ {\frac {A} {N-1}}s \sum_{r=0}^\infty |a_r| \frac 1 {(N-1)^r} 
\big ( r s (N-1)^{r-1}+ \frac {r(r-1)}2 (N-s-1)^{r-1} \big) \nn \\
&&\leq   8e^ {\frac {A} {N-1}}  \frac {s^2}{N-1} \sum_{r=0}^\infty |a_r| r^2 \leq 8Ae^ {\frac {A} {N-1}}  \frac {s^2}{N-1},
\eea
where we used that the sum in the second inequality is bounded by $A$ due to \eqref{hypo} and the fact that $r^2 \leq 8^r$.

To estimate $E_s^3$ we have
$$
E_s^3=E_s^{3,1} +E_s^{3,2}
$$
where
\be
\label{E31}
E^{3,1}_s(t)=- T_1    \sum_{r=0}^\infty  a_r \, C^N_{s,s+r+1} \, f^N_{s+r+1}
 \ee
and
\be
\label{E32}
E^{3,2}_s(t)= T_2    \sum_{r=0}^\infty  a_r \, C_{s,s+r+1} \, f^N_{s+r+1}
\ee
where
$$
T_1 := 1- (N-s) \a_N
$$
and
$$
T_2:=\frac {(N-s-1) \dots (N-s-r)}{(N-1)^r} -1.
$$

Moreover
\bea
T_1=  && 1- \frac {N-s}{(N-1) (1+e_K(N))} \nn \\
&&=  \frac {s-1}{(N-1) (1+e_K(N))} +  \frac {e_K(N)}{ (1+e_k(N))} \nn.
\eea

Therefore since $A>1$, using \eqref{eK} and \eqref{esta}, we obtain
\bea
\label{insert1}
|T_1|  && \leq  \frac {s-1}{(N-1)} 4e^{|e_K(N)| }+   4 \frac A {N-1} e^{|e_K(N)|} \nn \\
&& \leq  4 e^{\frac {A}{N-1}} (\frac {s-1}{N-1} + \frac {A}{N-1}) \nn \\
&&  \leq 8Ae^ {\frac {A} {N-1}}  \frac {s}{N-1} .
\eea

Finally
\bea
\label{insert2}
|T_2| \leq |\frac {(N-s-1) \dots (N-s-r)}{(N-1)^r} -1| && \leq |\frac {(N-s-r)^r-(N-1)^r} {(N-1)^r } |\nn \\
\leq \frac {r (s+r) (N-1)^{r-1}} {(N-1)^r} \leq \frac {2r^2s} {N-1} . 
\eea

As matter of facts by using \eqref {boundC} we conclude that
\be
\label {estE3}
\| E^3 _s(t) \|_{L^1} \leq   10 A^2e^ {\frac {A} {N-1}}  \frac {s^2}{N-1} .
\ee

Summarizing:

{\bf Proposition 1}

{\it We have 
\be
\label{prop}
\| E _s(t) \|_{L^1} \leq 22 A^2 e^ {\frac {A} {N-1}}  \frac {s^2}{N-1}  .
\ee
}
\bigskip

\section{Convergence}  \label{sec:2}
\setcounter{equation}{0}    
\def\theequation{6.\arabic{equation}}

In this section we  estimate the quantity
\be
\label {delta}
\D^N_s (t)=f^N_s(t) - f_s(t)
\ee
where $f^N_s(t) $ and $f_s(t)$ solve the initial value problems \eqref {hie3} and \eqref{hielim} respectively.
Taking the difference between \eqref {hie3} and \eqref{hielim}, we have 
\bea
\label{hiedelta}
(\pa_t +\sum\limits_{i=1}^{s} &&v_i \cdot \nabla_{x_i} )\D_s^N(t)=-s \D_s^N (t)+ E_s(t) \\
&& +    \sum_{r=0}^\infty  a_r \, C_{s,s+r+1} \, \D^N_{s+r+1} , 
 \nn
\eea 
with initial datum
$$
\D^N_s (0) = f_0^{\otimes s} {\bf 1}_{\{ s > N \}},
$$
where $C$ and $E$ are given by \eqref{Clim} and \eqref{E}.

We define the operator  $ S_j(t): L^1( X_j,V_j) \to L^1( X_j,V_j)$ by
\be
\label{S}
(S_j(t) f_j) (X_j,V_j) = e^{-jt} f_j(X_j -V_j t, V_j)
\ee
and notice that 
\be
\label{S}
\| S_j(t) \|_{L^1 \to L^1} \leq 1,
\ee
where $ \| \cdot \|_{L^1 \to L^1}$ denotes the operator norm.

We can express \eqref{hiedelta} in integral form
\bea
\label{hiedeltaint}
\D_j^N(t)=&&S_j(t-t_1) \D_j^N(t_1)+ \\
&& \int_{t_1}^t d \tau S_j (t-\tau)  \sum_{r=0}^\infty  a_r \, C_{j,j+r+1} \, \D^N_{j+r+1}(\tau)  \nn \\
&&+ \int_{t_1}^t d \tau S_j (t-\tau) E_j(\tau) .   \nn 
\eea 
for any $t_1 \in [0,t)$.

Therefore we can represent the solution $\D_j^N(t)$ as a series expansion in terms of the initial datum  $\D_j^N(t_1)$
and $E_j(s)$. 
To this end we define the operator  ${\cal T}_n (t,t_1)$ by recurrence.  For any sequence 
$F=\{ F_j \}_{j=1}^\infty $, $F_j \in L^1(X_j,V_j)$, set:
$$
({\cal T}_0 (t,t_1) F)_j =S_j (t-t_1) F_j
$$
and
$$
({\cal T}_n (t,t_1) F)_j = \int_{t_1}^t d \tau S_j (t-\tau)  \sum_{r=0}^\infty  a_r  \, C_{j,j+r+1} \, ({\cal T}_{n-1}  (\tau,t_1) F)_{j+r+1}.
$$
Therefore, denoting by $ \D^N$ and $E$ the sequences  $\{ \D_j \}_{j=1}^\infty $ and $\{ E_j \}_{j=1}^\infty $ respectively, by a standard computation we have
\be
\label{sN}
\D^N(t)=\sum_{n \geq 0}  {\cal T}_n (t,t_1) \D^N (t_1)+ \sum_{n \geq 1}\int _{t_1}^t ds {\cal T}_n (t,s ) E(s).
\ee

We are now in position to establish the  main result of the present paper

\medskip

{\bf Theorem 1} {\it For any  $T>0$ and $\a >\log2$,  there exists $N(T,\a)$  such that for any $t \in (0,T)  $, any $j \in \N$ and for  any $N>N(T,\a) $,
we have
\be
\label{th}
\| \D_j^N (t) \|_{L^1}  \leq 2^j \big(\frac 1 {N-1} \big )^{e^{-\alpha  (8At+1)} } .
\ee
}

{\bf Remark}

{\it Note that according to \eqref{th} the quality of the order of convergence rate deteriorates with increasing time. Note also that the magnitude of the error increases exponentially with the order $j$ of the marginals. In paticular if $j$ increases with $N$ too fast, correlations are persistent in the limit $N \to \infty$. }

\bigskip

{\bf Proof.}

The proof follows two steps. First we estimate ${\cal T}_n (t,t_1)$, and hence $\D^N(t)$  for a short time interval  $\d=t-t_1$. 
Then we split the time interval $(0,t)$ into $m$ intervals of length $\d$, with $\d$ small enough, to obtain the result inductively. 

\subsection{Short time estimate}  

We first observe, using \eqref{S},  that
\be
\label{estT}
\| ({\cal T}_n (t,t_1) F)_j \|_{L^1} \leq j  \sum_{r=0}^\infty  |a_r|  \int_{t_1}^t d \tau   \|  ({\cal T}_{n-1}  (\tau,t_1) F)_{j+r+1} \|_{L^1}.
\ee
Iterating this inequality and using, for $t>t_1$
$$
\int_{t_1}^t d\tau_1 \int_{t_1}^{\tau_1} d\tau_2 \cdots \int_{t_1}^{\tau_{n-1}} d\tau_n = \frac {(t-t_1)^n} {n!},
$$
we obtain, for any $F= \{ F_j \}_{j=1}^\infty $, setting $\d =\frac 1{8A}$ and  $R=\sum_{i=1}^{n-1}  r_i$, 
\bea
\label{st}
\hspace{-1cm}
\|  ({\cal T}_n (t,t-\d) F)_j \|_{L^1}& \leq &
 \frac {\d^n }{n!} \sum_{r_1 \dots r_n} |a_{r_1} | \cdots |a_{r_n} |  \nn \\
&& j(j+r_1+1) \cdots  (j+R+n-1)\| F_{j+R+n} \|_{L^1} \nn \\
&& \leq \sum_{r_1 \dots r_n} |a_{r_1} | \cdots |a_{r_n} | 2^{j+R-1} (2 \d)^n \| F_{j+R+n} \|_{L^1}.
\eea
In the last step, we used that
\bea
\frac { j(j+r_1+1) \cdots (j+R+n-1)}{n!} && \leq \frac { (j+R) (j+R+1) \cdots (j+R+n-1)}{n!} \nn \\
&& \leq \frac { (j+R+n-1)!}{n! (j+R-1)!} \leq 2^{j+R+n-1}. \nn
\eea
Applying \eqref {st} when $F=E$  with $t-\d$ replaced by $s$, we get, by Proposition 1,
\bea
\label{stE}
\int _{t-\d}^t ds && \|({\cal T}_n (t,s ) E(s))_j \|_{L^1} \leq  \nn \\
&& \leq C A^2 e^ {\frac {A} {N-1}}   \d \sum_{r_1 \dots r_n} |a_{r_1} | \cdots |a_{r_n} | 2^{j+R-1} (2 \d)^n \frac { (j+R+n)^2 }{N-1}.
\eea
where from now on $C$ will denote a positive numerical constant.
Moreover
$$
(j+R+n)^2 < 3n^2 +3j^2+3R^2
$$
so that
\bea
2^{j-1}  \sum_{r_1 \dots r_n} |a_{r_1} | \cdots  && |a_{r_n} | 2^{R} (R+j+n)^2 \leq C 2^{j-1} A^n (1+j^2+n^2) \nn \\
&& \leq C  A^2 e^ {\frac {A} {N-1}}  2^j A^n j^2 n^2
\eea
Here and in the sequel we use systematically 
$$
 \sum_{r_1 \dots r_n} |a_{r_1} | \cdots |a_{r_n}| \,\,\, 8^{(r_1+r_2+ \cdots r_n)} \leq A^n.
$$
Finally summing over $n$, using that, for $x \in (0,1)$
$$
\sum_{n=1}^\infty n^2 x^n = \frac {3 x} {1-x},
$$
we conclude that, recalling that $\d=\frac 1{8A}$
\bea
\label{stE1}
\sum_{n \geq 1} \int _{t-\d}^t ds && \|({\cal T}_n (t,s ) E(s))_j \|_{L^1}  \leq \nn \\
&& C A^2 e^ {\frac {A} {N-1}}  \d 2^j j^2 \frac {6A\d}{(1-2A\d)}  \frac 1 { N-1} = C(A) 2^j j^2 \frac 1 {N-1},
\eea
where $C(A)$ is a constant depending only on $A$.

\subsection{Iteration} 
Given an arbitrary $t>0$ we split the time interval $(0,t)$ in intervals $(k\d, (k+1)\d)$ $k=1 \cdots m$ where $m$ is an integer for which $t \in ( (m-1)\d, m\d]$.

Denoting
\be
\label{ind}
D_j(k) = \sup_{ s \in ((k-1)\d, k\d )} \| \D^N_j (s) \|_{L^1} , \,\, k=1\cdots m, 
\ee
with $D_j(0)= \D^N_j (0)=-f_0^{\otimes j} {\bf 1}_ {j >N} $,
we assume inductively that,  for $\a$ to be fixed later
\be
\label{k-1}
D_j(k-1) \leq 2^j \varphi (k-1,N)  \quad  \text {with} \quad \varphi (k,N)=\frac 1 { (N-1)^{e^{-\a k}}}.
\ee
We want to prove that the same holds for $k$, namely 
\be
\label{k}
D_j(k) \leq 2^j \varphi (k,N).
\ee

Note that the proof of the theorem is easily achieved once \eqref{k} is proven.

\eqref {k} is trivially true for $k=0$ since
$$
D_j(0) \leq 2^j 2^{-N}.
$$

Assuming \eqref {k-1} and applying  \eqref {st}  and \eqref {stE1} to \eqref{sN}, with $t \in ((k-1) \d, k \d)$, $t_1=(k-1)\d$ 
and $F=\D^N ((k-1) \d)$,
we have
\bea
\label {D}
D_j(k) && \leq \sum_{n \geq 0} \sum_{r_1 \dots r_n} |a_{r_1} | \cdots |a_{r_n} | 2^{j+R-1} (2 \d)^n 2^{j+R+n} \varphi (k-1,N)
\nn \\
&& +j^2 2^j \frac {C(A)}{N-1}
\eea

Now observe that $D_j(k) \leq 2$ so that \eqref{k} holds true whenever $j$ is so large to satisfy
\be
\label{jlarge}
2^j \varphi (k,N) >2.
\ee
Otherwise
\be
\label{jsmall}
2^j \leq \frac 2 { \varphi (k,N) }
\ee
or, equivalently
\be
\label{jsmall1}
j \leq 1+ \frac {e^{-\a k}} {\log 2} \log (N-1).
\ee
Using \eqref{jsmall1}, we control the second term in the right hand side of \eqref{D} by
$$
 2^j \varphi (k,N) \big\{ C(A) ( 1+ \frac {e^{-\a k}} {\log 2} \log (N-1))^2  \big( \frac 1 {N-1} \big)^{1-e^{-\a k}} \big \}.
$$
Now it is clear that 
$$
\{ \cdots \} \leq \frac12
$$
provided that $N$ is sufficiently large depending on $\a$ and $k$ (and hence on $t$).

On the other hand the first term in the right hand side of \eqref {D} is bounded by (using  \eqref{jsmall})
\bea
\sum_{n \geq 0} A^n 2^j 2^{j-1}(4 \d)^n \varphi (k-1,N) \leq &&2^j   \frac 1 { 1-4A\d} \varphi (k-1,N) (N-1)^{e^{-\alpha k} } \nn \\
&& \leq \frac 12  2^j  \varphi (k,N).
\eea
The last step follows from the fact that
$$
(N-1)^{e^{-\alpha k}} ( \frac 1 {N-1})^{e^{-\a (k-1)}} = (\frac 1 {N-1})^{e^{-\a k}} (\frac 1 {N-1})^{e^{-\a k} (e^{\a}-2)}
\leq \frac 12 (\frac 1 {N-1})^{e^{-\a k}}
$$
for $\a>\log 2$ and $N$  sufficiently large.

This concludes the proof.  \qed

\end{document}